\documentclass[twocolumn,aip,jcp,reprint,superscriptaddress,showpacs,floatfix]{revtex4-1}
\usepackage[latin9]{inputenc}
\usepackage{bm}
\usepackage{amsmath}
\usepackage{amssymb}
\usepackage{graphicx}
\usepackage{color}
\usepackage{CJK}



\begin{document}
\title{Effective Entropy Production and Thermodynamic Uncertainty Relation
of Active Brownian Particles}
\author{Zhiyu Cao}
\author{Jie Su}
\author{Huijun Jiang}
\author{Zhonghuai Hou}
\thanks{E-mail: hzhlj@ustc.edu.cn}
\affiliation{Department of Chemical Physics \& Hefei National Laboratory for Physical
Sciences at Microscales, iChEM, University of Science and Technology
of China, Hefei, Anhui 230026, China}
\date{\today}
\begin{abstract}
Understanding stochastic thermodynamics of active Brownian particles
(ABPs) system has been an important topic in very recent years. However,
thermodynamic uncertainty relation (TUR), a general inequality describing
how the precision of an arbitrary observable current is constraint
by energy dissipation, has not been fully studied for many-body level. Here, we
address such an issue in a general model of active Brownian particles
system by introducing an effective Fokker-Planck equation, which
allows us to identify a generalized entropy production only by tracking the stochastic trajectory of particles' position, wherein an activity and configuration dependent
diffusion coefficient comes into play an important role. Within this framework,
we are able to analyze the entropic bound as well as TUR associated
with any generalized currents in the systems. Furthermore, the effective
entropy production has been found to be a reliable measure to quantify
the dynamical irreversibility, capturing the interface and defects of motility induced phase separation
(MIPS). We expect the new conceptual quantities proposed here to be
broadly used in the context of active matter.
\end{abstract}
\maketitle

\section{Introduction}

Over the past two decades, stochastic thermodynamics has gained extensive
attention for describing nonequilibrium thermodynamics of mesoscopic
systems\cite{sekimoto2010stochastic,seifert2005entropy,seifert2008stochastic,jarzynski2011equalities,seifert2012stochastic}.
Due to the small size of such systems, fluctuations are significant,
so that thermodynamic quantities become stochastic variables. This
observation allows ones to generalize laws of thermodynamics at single
trajectory level\cite{gomez2011fluctuations,koski2015chip,martinez2017colloidal},
which leads to the study of stochastic energetics and fluctuation
theorems (FT). In particular, an important universal inequality between
the fluctuation in currents and thermodynamic cost, the thermodynamic
uncertainty relation (TUR), has been discovered\cite{barato2015thermodynamic,gingrich2016dissipation,pigolotti2017generic,pietzonka2016universal,polettini2016tightening,koyuk2018generalization,horowitz2017proof,dechant2018current,hasegawa2018thermodynamics,van2019uncertainty,potts2019thermodynamic,koyuk2019operationally,marsland2019thermodynamic}.
Specifically, TURs constrain the Fano factor of an arbitrary observable
current by the total entropy production, presenting a trade-off relation between precision and dissipation, and provide an alternative method to obtain a lower
bound on the entropy production. Moreover, TURs make an irreplaceable
contribution to our understanding of non-equilibrium phenomena (e.g.,
work extraction under measurement and feedback\cite{van2019uncertainty,potts2019thermodynamic,sagawa2012nonequilibrium}
and biological clocks\cite{barato2016cost}) which can provide more
detailed information about the systems than the second law. TUR was
first proposed for biological processes by Barato and Seifert\cite{barato2015thermodynamic}
and then extended to many other situations, such as diffusion process\cite{pigolotti2017generic,polettini2016tightening},
finite-time generalization \cite{pigolotti2017generic,pietzonka2016universal,dechant2018current},
periodically driven systems \cite{koyuk2018generalization,koyuk2019operationally},
biological oscillators \cite{hasegawa2018thermodynamics,marsland2019thermodynamic,nguyen2018phase,cao2020design},
time-delayed systems \cite{van2018thermodynamic}.

Very recently, the frontiers of stochastic thermodynamics have shifted
to the active particle systems. Active particles form a class of nonequilibrium
systems which have the ability to perform directional motion
through self-propulsion by consuming energy from the environment\cite{vicsek2012collective,marchetti2013hydrodynamics,bechinger2016active,fily2012athermal,buttinoni2013dynamical,giomi2008complex,peruani2006nonequilibrium,redner2013structure,paoluzzi2016critical}.
Study of active particles encompasses a wide variety of biological and
soft matter systems, such as schools of fish\cite{vicsek2012collective},
biological microorganisms\cite{Czirok2001theory,Sokolov2009enhanced,Frydel2022the,Felderhof2017swimming,Jin2020influence}
and colloidal particles\cite{Ivanov2018anomalous, Shin2020Diffusiophoretic, Fujitani2022Diffusiophoresis}. Such systems can
usually show a range of typical collective behaviors including phase
separation\cite{stenhammar2015activity,Stenhammare2016light,pu2017reentrant},
turbulence\cite{Dunkel2013fluid,rank2021active,ghaemi2020passive}, self-assembly\cite{du2019self}. It has been reported that even the purely repulsive active
systems can yield a motility induced phase separation (MIPS) where
particles spontaneously separate into solid-like and gas phases\cite{fily2012athermal,redner2013structure},
which has attracted much attention. So far, the laws of thermodynamics,
the definition of related entropy production or the construction of
fluctuation theorem in active particle systems has been discussed in a few studies\cite{fodor2016far,mandal2017entropy,dabelow2019irreversibility,ganguly2013stochastic,speck2016stochastic,falasco2016exact,chaki2018entropy,chaki2019effects,caprini2019entropy,speck2018active,maggi2014generalized,nardini2017entropy,shankar2018hidden,busiello2019entropy,szamel2019stochastic,crosato2019irreversibility,cao2021designing}. Most studies require not only tracking the particles' position, but also full information including the self-propulsion\cite{ganguly2013stochastic,szamel2019stochastic,crosato2019irreversibility,cao2021designing}, which is a great challenge in experiments\cite{hoang2018experimental}. Therefore, how to understand the behaviors of many-body active particle systems from a thermodynamics perspective in most practical scenarios when only partial information is available, such as the establishment of TUR, is a great challenge.

In this article, based on the approximation of mapping the active particles
system to an ``effective equilibrium'' one\cite{farage2015effective,Wittmann2017effective,fox1986functional,fox1986uniform}, we have introduced a generalized entropy production
(EP) $S_{g}$, which only needs to track the position of particles. We compare the proposed EP with definitions in other studies and demonstrate the hierarchical order between them by analyzing the degree of coarse-graining. Direct simulations help us to identify the generalized EP at each
point in the phase diagram. Detailed analysis of the spatial distributions
of the proposed quantity allows us to identify the interface and defects of MIPS, which means that it can unambiguously be utilized to measure the
dynamic irreversibility on a macroscopic scale. Furthermore,
the entropic bounds\cite{dechant2018entropic,Shiraishi2016universal}
and generalized TURs\cite{barato2015thermodynamic,gingrich2016dissipation,pigolotti2017generic,pietzonka2016universal,polettini2016tightening,koyuk2018generalization,horowitz2017proof,dechant2018current,hasegawa2018thermodynamics,cao2020design,van2019uncertainty,van2018thermodynamic,potts2019thermodynamic,koyuk2019operationally,marsland2019thermodynamic,cao2021designing}
of active systems based on the demonstrated approximation, providing a
convenient tool for entropy production inference\cite{seifert19from}.
\section{Model}

We consider a homogeneous system of $N$ active Brownian particles
with spatial coordinates $\bm{x}(t)=\{\bm{x}_{i}(t)\}$, self-propelling with constant
velocities $v_{0}$ along its direction of orientations $\bm{n}_{i}=(\cos\theta_{i},\sin\theta_{i})$ for $i$-th particle.
Assume that the particles move in a viscous medium and hydrodynamic
interactions are neglected, the resulting governing equations are:

\begin{equation}
\dot{\bm{x}}(t)=\mu\bm{F}(\bm{x})+v_{0}\bm{n}(t)+\bm{\xi}(t),\label{eq:ABP1}
\end{equation}

\begin{equation}
\dot{\theta}_{i}(t)=\zeta_{i}(t)\label{eq:ABP2}
\end{equation}
Here, $\bm{F}=-\nabla_{\bm{x}}U$ is the mechanical force generated
from the total interactions $U(\bm{x})$ and $\mu$ is the mobility.
The stochastic terms $\bm{\xi}_{i}(t)$ and $\zeta_{i}(t)$ are Gaussian
white noises with correlations $\langle\xi_{i}(t)\xi_{j}(s)\rangle=2D_{t}\delta_{ij}\delta(t-s)$
and $\langle\zeta_{i}(t)\zeta_{j}(s)\rangle=2D_{r}\delta_{ij}\delta(t-s)$.
The translational diffusion coefficient $D_{t}$ satisfies $D_{t}=\mu k_{B}T$
with $k_{B}$ the Boltzmann constant (which is set to be $1$ throughout
the paper) and $T$ the ambient temperature. The rotational diffusion
coefficient $D_{r}$ relates to persistent time as $\tau_{p}=(2D_{r})^{-1}$.
To derive an explicit TUR for ABPs system, we now introduce
a coarse-grained active Ornstein-Uhlenbeck process with thermal noise (AOU-T) as a direct mapping model, which reads
\begin{equation}
\dot{\bm{x}}(t)=\mu\bm{F}(\bm{x})+\bm{\xi}(t)+\bm{\eta}^{A}(t).\label{eq:AOUT}
\end{equation}
Here, the active term $\bm{\eta}^{A}(t)$ represents the OU active
components with zero mean and time correlation
\begin{equation}
\langle\eta_{i}^{A}(t)\eta_{j}^{A}(s)\rangle=\frac{v_{0}^{2}}{3}e^{-|t-s|/\tau_{p}}\delta_{ij},\label{eq:COU}
\end{equation}
In the limit $\tau_{p}\rightarrow0$, the time
correlation becomes $\langle\eta_{i}(t)\eta_{j}(s)\rangle=2D_{a}\delta_{ij}\delta(t-s)$
with $D_{a}=v_{0}^{2}\tau_{p}/3$, i.e, the system reduces to an equilibrium
one with effective diffusion coefficient $D_{t}+D_{a}$.

Over the past decade, study of ABPs is a very
hot topic and has gained extensive research attention. In particular,
stochastic thermodynamics of ABPs has become a frontier area very
recently. The main motivation of the present work is to address a TUR for many-body ABPs system. Following the scheme proposed by Seifert\cite{seifert2005entropy},
one can define the EP of the system along a given
stochastic trajectory $\chi(t)=\{\bm{x}(t)|_{t=0}^{t=t_{f}}\}$ as
$S_{sys}(t)=-\ln P(\bm{x},t)$, where $P(\bm{x},t)$ is the configurational
probability distribution for the state variable to take the value
$\bm{x}$ at time $t$. In order to establish the framework of stochastic thermodynamics of many-body active systems, we need to obtain the Fokker-Planck equation (FPE) which governs the evolution of probability distribution $P(\bm{x},t)$\cite{farage2015effective}. To proceed,
we adopt the Fox method\cite{fox1986functional,fox1986uniform}
to get an effective FPE which can best approximate the process of
physical interests and make accurate predictions\cite{sharma2017escape,scacchi2019escape,scacchi2018mean},
which reads
\begin{equation}
\partial_{t}P(\bm{x},t)\approx -\sum_{i=1}^{N}\partial_{x_{i}}J_{i}(\bm{x},t).\label{eq:FPE}
\end{equation}
The probability current is given by
\begin{equation}
J_{i}(\bm{x},t)=D_{i}(\bm{x})\beta F_{i}^{eff}(\bm{x})P(\bm{x},t)-D_{i}(\bm{x})\partial_{x_{i}}P(\bm{x},t),\label{eq:J}
\end{equation}
where $\beta=1/T$, and $D_{i}(\bm{x})$ denotes a configuration-dependent diffusion coefficient given by
\begin{equation}
D_{i}(\bm{x})=D_{t}+D_{a}\left[1-\beta\tau\partial_{x_{i}}F_{i}\left(\bm{x}\right)\right]^{-1}\label{eq:Deff}
\end{equation}
with $\tau=\tau_{p}D_{t}/d^{2}$ a dimensionless persistence time
and $d$ the typical diameter of a particle. $F_{i}^{eff}(\bm{x})=D_{i}^{-1}(\bm{x})[D_{t}F_{i}(\bm{x})-T\partial_{x_{i}}D_{i}(\bm{x})]$
gives the effective force exerting on the $i$-th particle. For a
passive system in the absence of $\bm{\eta}^{A}(t)$, $D_{i}\left(\bm{x}\right)=D_{t}$
and $F_{i}^{eff}\left(\bm{x}\right)=F_{i}\left(\bm{x}\right)$, while
in the limit $\tau\rightarrow0$, $D_{i}\left(\bm{x}\right)=D_{t}+D_{a}$
and $F_{i}^{eff}\left(\bm{x}\right)=D_{t}F_{i}\left(\bm{x}\right)/(D_{t}+D_{a})$.
The Fox method is an approximation which is valid in lower powers of the
persistence time $\tau_{p}$, as shown in the original paper\cite{fox1986functional,fox1986uniform}.
Nevertheless, it may go beyond this by including
contributions to higher orders in $\tau_{p}$\cite{farage2015effective,Wittmann2017effective}.
Indeed, detailed studies of different systems demonstrate the
validity of Fox approximation over a large range of $\tau_{p}$ values\cite{feng2017mode}.
Thus, one may expect that Fox approximation could be applied in a
certain range of $\tau_{p}$, the exact values of which may be system-dependent. In the current system of active particles, there still
exists another condition for the Fox approximation to be valid, i.e.,
$1-\beta\tau\partial_{i}F_{i}\left(\bm{x}\right)>0$, such that $D_{i}\left(\bm{x}\right)$
is positive in the entire area. Thus, the range of accessible $\tau_{p}$
values depends upon the specific form of the bare interaction potential.

\section{Effective entropy production}

Based on the Fox approximation, the AOU-T equation Eq.(\ref{eq:AOUT}) corresponds to an equivalent Langevin
equation. Within this framework, $\bm{F}^{eff}(\bm{x})$ represents
the effective interparticle force done on the particle which is related
to the heat flux of the system. According to Sekimoto's suggestion\cite{sekimoto2010stochastic},
we can define a generalized heat dissipation in the medium along a
stochastic path $\chi(t)$ as
\begin{equation}
\Sigma_{m}[\chi]=\int_{0}^{t_{f}}\bm{F}^{eff}(\bm{x})^{\text{T}}\circ\dot{\bm{x}}dt,\label{eq:Sm}
\end{equation}
where ``$\circ$'' stands for the Stratonovich product and the superscript
`T' means transposition. Since the effective force $\bm{F}^{eff}(\bm{x})$
is the total force done on the system including the effect of activity,
the generalized heat dissipation will recover to the normal heat dissipation
$\int_{0}^{t_{f}}\bm{F}(\bm{x})^{\text{T}}\circ\dot{\bm{x}}dt$ in
passive systems\cite{seifert2005entropy}. The difference between
these two types of heat dissipation is the activity-induced extra
entropy flux.

According to the Eq.(\ref{eq:J}), the change rate of the system
entropy is
\begin{align}
\dot{S}_{sys}\left(t\right) & =-\partial_{t}\ln P(\bm{x},t)\nonumber \\
 & =-\frac{1}{P(\bm{x},t)}[\frac{\partial P(\bm{x},t)}{\partial t}+\sum_{i=1}^{N}\partial_{x_{i}}P(\bm{x},t)|_{\bm{x}(t)}\dot{x}_{i}]\nonumber \\
 & =-\frac{1}{P(\bm{x},t)}[\frac{\partial P(\bm{x},t)}{\partial t}-\sum_{i=1}^{N}\frac{J_{i}(\bm{x},t)}{D_{i}(\bm{x})}|_{\bm{x}(t)}\dot{x}_{i}]\nonumber \\
 & \ -\beta\bm{F}^{eff}(\bm{x})^{\text{T}}\dot{\bm{x}}.\label{eq:dS}
\end{align}
Clearly, the final term in the third equality is related to the generalized
heat dissipation in Eq.(\ref{eq:Sm}), i.e., $\beta\bm{F}^{eff}(\bm{x})^{\text{T}}\circ\dot{\bm{x}}=\dot{\Sigma}_{m}/T$.
Then, Eq.(\ref{eq:dS}) can be rewritten as a balance equation for
the trajectory-dependent total EP $\dot{S}_{g}=\dot{\Sigma}_{m}/T+\dot{S}_{sys}$,
\begin{equation}
\dot{S}_{g}\left(t\right)=-\frac{\partial_{t}P(\bm{x},t)}{P\left(x,t\right)}|_{\bm{x}(t)}-\sum_{i=1}^{N}\frac{J_{i}(\bm{x},t)}{D_{i}(\bm{x})P\left(x,t\right)}|_{\bm{x}(t)}\dot{x}_{i}.\label{eq:dSg}
\end{equation}
By averaging over the path ensemble, we can obtain the following equation
\begin{equation}
\langle\dot{S}_{g}\left(t\right)\rangle\approx\sum_{i=1}^{N}\int\frac{J_{i}^{2}(\bm{x},t)}{D_{i}(\bm{x})P(\bm{x},t)}d\bm{x}\geq0,\label{eq:2nd}
\end{equation}
where $\int d\bm{x}\partial_{t}P(\bm{x},t)=0$ and $\langle\dot{x}_{i}|\bm{x},t\rangle\approx J_{i}(\bm{x},t)/P(\bm{x},t)$
have been used. All the information about particle activity
is contained in $D_{i}\left(\bm{x}\right)$ and $J_{i}\left(\bm{x}\right)$.
The second law Eq.(\ref{eq:2nd}) ensures that the averaged total
EP must increase with time. The equality holds if all the currents
$J_{i}\left(x\right)$ vanish, i.e., the system can be exactly mapped to an
equivalent equilibrium system.

In general, the generalized EP $S_{g}$ along a stochastic trajectory is an apparent measure of the time-reversal symmetry broken of the system at the scale of observed trajectories. In the current work, the proposed EP can be obtained simply by tracking the trajectory $\chi(t)=\{\bm{x}(t)|_{t=0}^{t=t_{f}}\}$
for particle positions $\bm{x}(t)$. To this end, two steps have been
used. Firstly, the orientational degree of freedom $\bm{n}(t)$ has
been eliminated and replaced by a colored noise within a mean-field level
of description. Secondly, the system with a non-Markovian colored
noise is approximated to an ``effective equilibrium'' one on a coarse-grained
time scale via the Fox method. Therefore, the instantaneous entropy production rate
(EPR) $\dot{S}_{g}=\bm{F}^{eff}(\bm{x})^{\text{T}}\circ\dot{\bm{x}}/T+\frac{d}{dt}\ln P(\bm{x},t)$
defined in our work can be viewed as a ``coarse-grained'' measure
of dynamic irreversibility of the system (In steady states, $\dot{S}_{g}=\bm{F}^{eff}(\bm{x})^{\text{T}}\circ\dot{\bm{x}}/T$). Based on this effective
mapping, a clearcut TUR for the coarse-grained EPR can then be well established. Other frameworks for studying
EPR and related properties for the ABPs system have been proposed. Nevertheless, many-body TUR has not been addressed so far. In the following, we elucidate the the precise hierarchy of EPs.

Firstly, in Ref.\cite{szamel2019stochastic}, Szamel has proposed
an EPR $\dot{S}_{sz}=\dot{\Sigma}_{sz}/T+\frac{d}{dt}\ln P\left(\bm{x},\bm{n};t\right)$
for the active particles with the heat dissipation $\dot{\Sigma}_{sz}=\left(\bm{F}+\mu^{-1}v_{0}\bm{n}\right)^{\text{T}}\circ\dot{\bm{x}}$. $\dot{S}_{sz}$ was constructed at the full dynamics level described by Eqs.(\ref{eq:ABP1}) and (\ref{eq:ABP2}), including the information
of orientation trajectory with difficulty in tracking. As discussed above, the generalized heat
dissipation rate $\dot{S}_{g}$ proposed by us can be regarded as
a coarse-grained form of $\dot{S}_{sz}$, since the orientational
degree of freedom $\bm{n}(t)$ has been reduced and the memory effects
have been coarse-grained. The difference between $\dot{S}_{g}$ and
$\dot{S}_{sz}$ is commonly referred to as ``hidden EPR'', which can be identified as the loss of information quantifying the
correlation between particle trajectory and the active term\cite{esposito2012stochastic,celani2012anomalous,kawaguchi2013fluctuation,chun2015hidden,wang2016entropy,shankar2018hidden,dabelow2019irreversibility,busiello2019entropy,crosato2019irreversibility}.

Secondly, based on the AOU-T model Eq.(\ref{eq:AOUT}), Debalow $et$
$al.$ has proposed an instantaneous EPR $\dot{S}_{da}=\bm{F}_{m}[\chi,t]^{\text{T}}\circ\dot{\bm{x}}/T$
(here $\bm{F}_{m}[\chi,t]$ is defined as the nonlocal ``memory forces'') which depends not only on $\bm{\ensuremath{x}}(t)$,
but also on the whole trajectory $\chi(t)=\{\bm{x}(t)|_{t=0}^{t=t_{f}}\}$
of the particles' position\cite{dabelow2019irreversibility}. In contrast, according to the definition of generalized
EPR in our work, $\dot{S}_{g}=\bm{F}^{eff}(\bm{x})^{\text{T}}\circ\dot{\bm{x}}/T+\frac{d}{dt}\ln P(\bm{x},t)$,
one can clearly find that it is only dependent on the current configuration
$\bm{x}(t)$ in the steady state, which is easily accessible in experiments. On the other hand, since AOU-T model is a coarse-grained form of Eqs.(\ref{eq:ABP1}) and (\ref{eq:ABP2}), we also have another EP hierarchy: $\dot{S}_{sz}>\dot{S}_{da}$.

Thirdly, in the literature, various continuous field theories based
on coarse-graining procedures have been proposed to capture the large
scale physics of active particles, such as ``Active Models'' A,
B, H\cite{Cates2019active}. In Ref.\cite{nardini2017entropy}, based on Active model B, Nardini $et$ $al.$ proposed an EPR $\dot{S}_{na}$ to quantify the dynamic irreversibility of the many-body active particle systems
at a macroscopic scale, even when phase separation happens. Therein,
the local steady-state EPR was defined as $\dot{S}_{na}=-\frac{1}{D}\left\langle \mu_{A}\dot{\phi}\right\rangle$,
where $\phi(\bm{x},t)$ denotes the fluctuating density field, $\mu_{A}$
is the additional contribution to the chemical potential due to the
effect of activity and $D$ is collective diffusivity. Due to its
field dependence, $\dot{S}_{na}$ is more ``coarse-grained'' than
our version $\dot{S}_{g}$.

At last, to highlight the effect of activity and effective interactions,
it would be instructive to consider a comparative case, where one
can treat the active system as another ``effective equilibrium''
system with a high effective temperature $T_{eff}=\mu^{-1}\left(D_{t}+D_{a}\right)$,
corresponding to the case in the limit $\tau_{p}\to0$. The corresponding
total EPR is given by $\dot{S}_{g}^{\left(1\right)}=\bm{F}\left(\bm{x}\right)^{\text{T}}\circ\dot{\bm{x}}/T_{eff}$.
Such scheme has been reported to establish the TUR for a single hot
Janus swimmer successfully\cite{falasco2016exact}. The difference
between $\langle\Delta S_{g}\rangle$ and $\langle\Delta S_{g}^{(1)}\rangle$,
treated as some kind of the ``hidden EP'' discussed above, is dominant
especially when the gradient of mechanical force $\partial_{x_{i}}F_{i}\left(\bm{x}\right)$
is significant.

Generally, due to the above discussion, the hierarchical structure of the
mentioned EPRs can be rationalized as $\dot{S}_{sz}>\dot{S}_{da}>\dot{S}_{g}>\dot{S}_{g}^{(1)}>\dot{S}_{na}$ by clarifying their corresponding degree of coarse-graining. Such a hierarchy has a similar counterpart in Maxwell's demon system\cite{barato2014unifying,horowitz2014second}.

In the following, we also present some discussion about the physical quantities by our formulation and the Harada-Sasa relation (HSR), which provides a useful tool to calculate the heat dissipation of the nonequilibrium systems\cite{harada2005equality}. Specifically, Harada and Sasa proposed an exact equality quantifying heat production in terms of the violation of the fluctuation dissipation relation (FDR), which has been developed for active particle systems. For instance, in Ref. \cite{nardini2017entropy}, Nardini $et$ $al.$ have derived a generalized HSR within a field-theoretical description of active matter. Chaki and Chakrabarti have utilized the HSR to calculate the heat dissipation of a colloidal particle immersed in an active bath \cite{chaki2019effects}. More recently, Jones $et$ $al.$ have analyzed the power dissipation of an active microswimmer ($Chlamydomonas$ $reinhardtii$) based on the HSR \cite{jones2021stochastic}. Besides, another meaningful application of the HSR is that the seemingly hidden entropy production can be partially probed from the violation spectrum of FDR \cite{wang2016entropy}. First of all, we state that both quantities can provide proper measures of deviation from equilibrium of many-body active particles. The difference is that in the experimental investigation, only the information of the spatial steady-state trajectories is required under our framework, while for HSR, the response spectrum and fluctuation spectrum of the system must be accessible. The amount of information contained in the violation spectrum determines how well the HSR probes the heat dissipation of the system. On the one hand, in small stochastic systems, it is a nontrivial task to directly measure the response functions, whereas details about spatial trajectories are easily observed. In fact, determining the response spectrum requires measuring each frequency separately, which must cover the high frequency region to ensure convergence of the integration in the HSR, substantially increasing the statistical effort \cite{Lander2012noninvasive}. On the other hand, to experimentally test the HSR, one needs to perturb the system. Conversely, our framework is noninvasive in experiments.

\section{Entropic bounds and thermodynamic uncertainty relation(TUR)}

According to Ref. \cite{dechant2018entropic,Shiraishi2016universal},
the entropic bounds on $\dot{S}_{g}$, stronger than the second law,
can be obtained by using the Cauchy-Schwarz inequality
\begin{equation}
\langle\dot{S_{g}}\rangle\leq\left\langle \bm{D}\cdot\left(\bm{F}^{eff}\right)^{2}\right\rangle .\label{eq:bound}
\end{equation}

This conclusion worthy of attention implies that the change rate of
the generalized EP is bounded by an activity and configuration-dependent
term. Actually, activity affects both sides of the equation, mainly
through the activity dependent diffusivity $\bm{D}\left(\bm{x}\right)$
and interaction $\bm{F}^{eff}(\bm{x})$.

Now, we turn to an important universal inequality between the fluctuations
in current and thermodynamic cost, the thermodynamic uncertainty relation
(TUR) \cite{barato2015thermodynamic,pietzonka2016universal,gingrich2016dissipation,horowitz2017proof,pigolotti2017generic,polettini2016tightening,van2018thermodynamic,potts2019thermodynamic,dechant2018current,hasegawa2018thermodynamics,van2019uncertainty,koyuk2018generalization,koyuk2019operationally,marsland2019thermodynamic,cao2020design,cao2021designing}.
To see the TUR of ABPs, we consider a generalized current $\Theta[\chi]$
along a single trajectory $\chi$ defined as \cite{van2018thermodynamic,dechant2018entropic}
\begin{equation}
\Theta[\chi]=\int\bm{\Lambda}(\bm{x})^{\text{T}}\circ\dot{\bm{x}}dt,\label{eq:CUR}
\end{equation}
where $\bm{\Lambda}(\bm{x})=\left(\Lambda_{1}\left(\bm{x}\right),\Lambda_{2}\left(\bm{x}\right),\dots,\Lambda_{N}\left(\bm{x}\right)\right)$
is a projection operator. Using different projection operator, one
can get different kinds of current, such as the moving distance of
particles or the EP in a time interval. The change rate of $\Theta$
can be written as
\begin{equation}
\langle\dot{\Theta}\rangle=\int\bm{\Lambda}(\bm{x})^{\text{T}}\bm{J}(\bm{x},t)d\bm{x}.\label{eq:DCUR}
\end{equation}
For instance, for the choice $\Lambda_{i}(\bm{x})=\delta_{ik}$ in
steady state, the generalized observable current is the drift velocity
of the $k-$th active particle and $\Psi_{i}(\bm{x})=\delta_{ik}D_{i}(\bm{x})$
only depends on the current configuration of the system.

Due to the markovity of effective dynamics, one can obtain\cite{dechant2018current}
\begin{equation}
\frac{Var[\Theta]}{\langle\Theta\rangle^{2}}\geq\frac{2}{\langle\Delta S_{g}\rangle},\label{eq:TUR-1}
\end{equation}
which serves as a TUR for currents based on information of particle-position
trajectories in the steady states. The effect of particle activity
is reflected in the activity-dependent total EP $\left\langle \Delta S_{g}\right\rangle $.
Mathematically, we have approximated the colored noise by a white
one, and obtained the effective Fokker-Planck equation which facilitates
following derivation. Only through this mapping, can we then set up
a TUR which can be numerically checked for real systems or even by
experiments. Several notable points are presented as follows. Firstly, in a steady
state, the change of system entropy vanishes and thus $\left\langle \Delta S_{g}\right\rangle =\left\langle \Sigma_{g}\right\rangle /T=T^{-1}\left\langle \int_{0}^{t_{f}}F^{eff}\left(\bm{x}\right)^{\text{T}}\circ\dot{\bm{x}}dt\right\rangle $.
Secondly, the TUR $\langle\Delta S_{g}\rangle\geq2\langle\Theta\rangle^{2}/Var[\Theta]$
gives the lower bound, and the entropic bound $\langle\dot{S_{g}}\rangle\leq\left\langle \bm{D}\cdot\left(\bm{F}^{eff}\right)^{2}\right\rangle $
provides the upper bound of the generalized EP. Further, as the generalized
EP will decrease after coarse-grained approximation, Eq.(\ref{eq:TUR-1}) can even be used to infer the exact EP including
the self-propulsion's contribution. Thirdly, in Ref.\cite{van2019uncertainty}, Tan Van Vu $et$ $al$.
indicated that the current fluctuation is constrained not only by
the entropy production but also by the average dynamical activity
in athermal AOU model, by mathematically mapping the system into an underdamped one. As with the framework in most studies, the dynamic irreversibility measure they define requires tracking the self-propelled velocities of active particles.

\section{Numerical Results of ABPs Model}

In this part, we have discovered the generalized EP as well as the
validity of entropic bound and TUR by direct numerical simulations.
We consider a system with $N$ disk-shaped particles in a two-dimensional
$xy$ plane. Here, the exclusive-volume pair potentials of ABPs are
modeled by the Weeks-Chandler-Anderson (WCA) potential: $U(r)=4\epsilon\left[\left(\frac{d}{r}\right)^{-12}-\left(\frac{d}{r}\right)^{-6}+\frac{1}{4}\right]$
for $r<2^{1/6},$ $\epsilon=k_{B}T$ and $U=0$. Here, $r=\left|\bm{x}_{1}-\bm{x}_{2}\right|$
is the particle separation, and $\epsilon$ is the interaction strength.
Proper values of parameters have been chosen to illustrate our main
results while $N=5120$ throughout the paper.
Here, we focus on the steady-state thermodynamics of the system, where
the system entropy change vanishes. We need to emphasize that the thermodynamic quantities are obtained from the trajectories generating
from Eqs.(\ref{eq:ABP1}) and (\ref{eq:ABP2}).

For a sufficiently large particle density $\phi$, our numerical results
show the MIPS: a coexistence between vapor
and dense at a critical self-propulsion velocity $v_{0}^{c}$($\phi$), which is phase density dependent.
To further demonstrate the phase separation, we introduce the local
order parameter with respect to particle $i$\cite{pu2017reentrant,crosato2019irreversibility}:

\begin{equation}
q^{6}(i)=\frac{1}{6}\left|\sum_{j\in N^{i}}\exp\left(i6\varphi_{ij}\right)\right|,
\end{equation}
where $N^{i}$ are the closest six neighboring particles of $i$ and
$\varphi_{ij}$ is the angle between the bond vector connecting particle
$i$ to $j$.

\begin{figure}
\begin{centering}
\includegraphics[width=1\columnwidth]{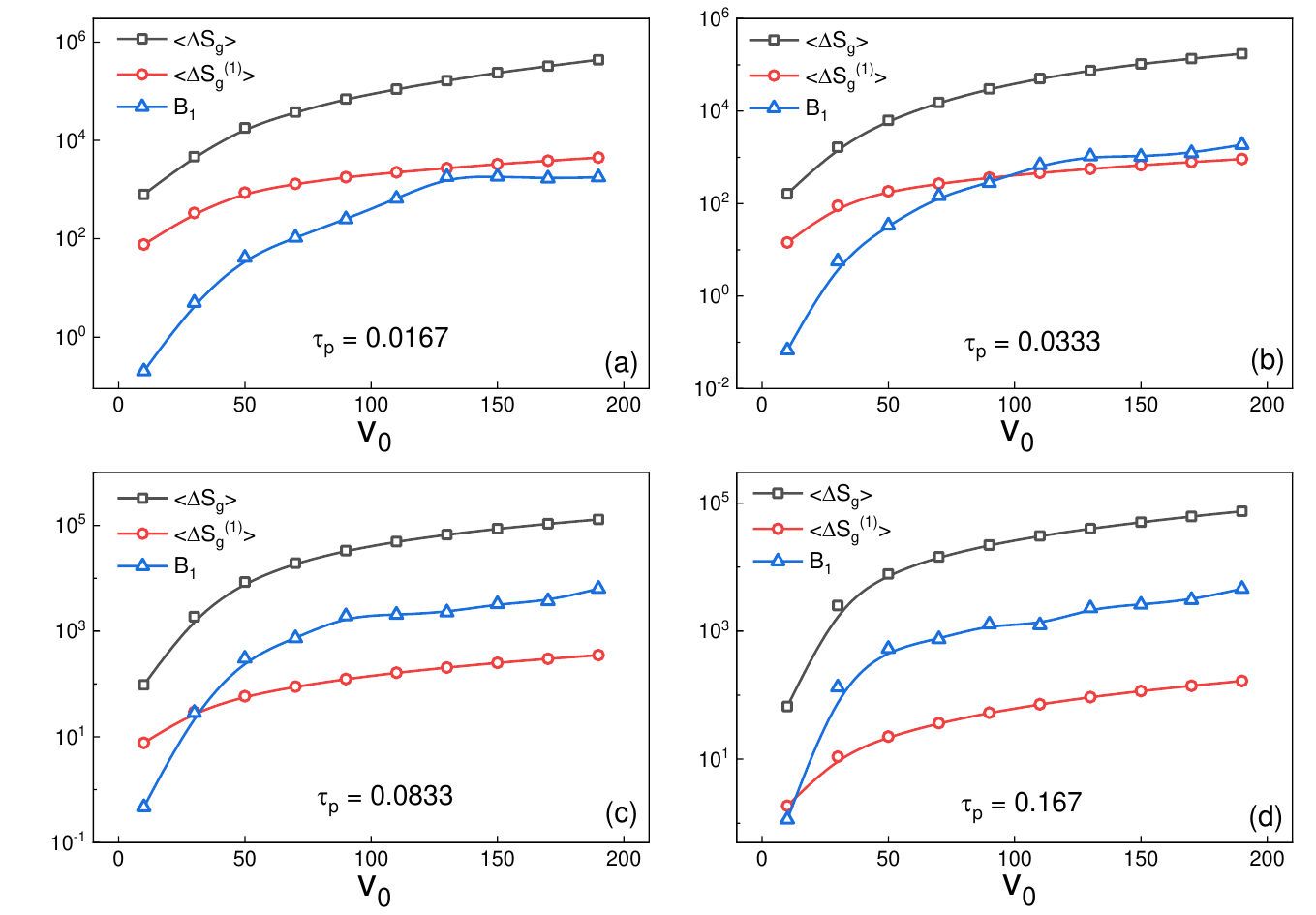}
\par\end{centering}
\caption{Validation of our key results, the TURs on the current $\Theta_{1}=S_{g}$.
Here, $B_{1}=B(\Theta_{1})=\frac{2\langle\Theta_{1}\rangle^{2}}{Var[\Theta_{1}]}$
measures the uncertainty of total entropy production and $v_{0}$
is the self-propelling velocity. We also plot the generalized entropy
production $\langle\Delta S_{g}\rangle$, and the alternative entropy
production $\langle\Delta S_{g}^{\left(1\right)}\rangle$ by treating
the active system as an equilibrium system with a high effective temperature
is also presented for comparison. Comparison between these quantities
from the simulations has been presented for different values of the
persistent time $\tau_{p}$: $\tau_{p}=0.0167$ (a), $\tau_{p}=0.0333$
(b), $\tau_{p}=0.0833$ (c), $\tau_{p}=0.167$ (d).}

\label{fig:1}
\end{figure}

\subsection{TUR and entropic bound}

To test the validity of TUR and entropic bound, we explore the ABPs
model over $v_{0}$, $\tau_{p}$ and $\phi$. In Fig.\ref{fig:1},
we begin by choosing the current $\Theta_{1}=S_{g}$ and plot the
generalized EP $\langle\Delta S_{g}\rangle$ and $\langle\Delta S_{g}^{(1)}\rangle$
with the TUR bound $B_{1}=B(\Theta_{1})=2\langle\Theta_{1}\rangle^{2}/Var[\Theta_{1}]$
over $\tau_{p}$ by varying $v_{0}$. Indeed, one can see that all
the data for $B_{1}$ lie below $\langle\Delta S_{g}\rangle$, demonstrating
the validity of our TUR Eq.(\ref{eq:TUR-1}). Nevertheless, if one use
$\langle\Delta S_{g}^{(1)}\rangle$ instead of $\langle\Delta S_{g}\rangle$,
obvious violation presents while increasing the persistent time $\tau_{p}$.
Therefore, the simulation results clearly vindicate that our method
to treat the many-body ABPs system correctly establishes the TUR in
a large range of persistence time $\tau_{p}$.

\begin{figure}
\begin{centering}
\includegraphics[width=1\columnwidth]{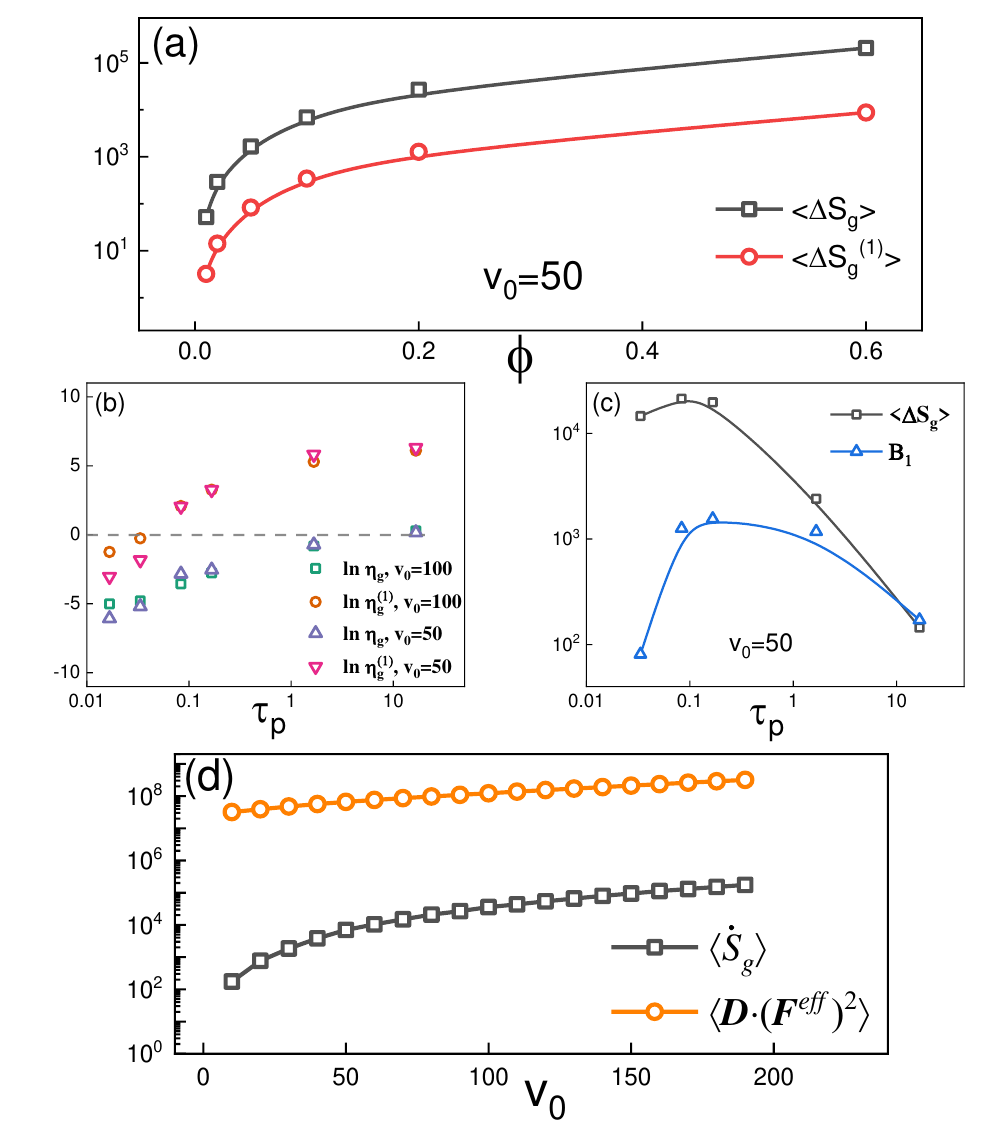}
\par\end{centering}
\caption{Further validation of the TURs and entropic bounds. (a) Comparison
between the TUR bound $B_{1}=B(\Theta_{1})=2\langle\Theta_{1}\rangle^{2}/Var[\Theta_{1}]$,
the generalized entropy production $\langle\Delta S_{g}\rangle$ and
the alternative entropy production $\langle\Delta S_{g}^{\left(1\right)}\rangle$
(normalized by the particles density) for different particle density
$\phi$. Here, we choose $\Theta_{1}=S_{g}$, $\tau_{p}=0.0167$ and
$v_{0}=50$. (b), (c) Numerical results of TURs of larger range $\tau_{p}$. (b) Log ratio of the TUR parameter $\eta_{g}=\frac{2\langle\Theta\rangle^{2}}{Var[\Theta]\langle\Delta S_{g}\rangle}$
and $\eta_{g}^{(1)}=\frac{2\langle\Theta\rangle^{2}}{Var[\Theta]\langle\Delta S_{g}^{(1)}\rangle}$
for $v_{0}=50$ and $v_{0}=100$. When $\ln\eta\protect\leq0$, our
TUR Eq.(\ref{eq:TUR-1}) is established, otherwise, the TUR is invalid.
(c) The non-monotonic persistent time dependence of the generalized
entropy production rate $S_{g}$ when $v_{0}=50$. (d) Numerical validation
of the entropic bound, Eq.(\ref{eq:bound}), for a larger range of
self-propelling velocity $v_{0}$. Here, we choose $\tau_{p}=0.0167$.}

\label{fig:2}
\end{figure}

In Fig.\ref{fig:2}(a), we also plot how the generalized EP, $\langle\Delta S_{g}\rangle$ and $\langle\Delta S_{g}^{(1)}\rangle$, and TUR
bound behave for different particle density $\phi$. We find that
the difference between $\langle\Delta S_{g}\rangle$ and $\langle\Delta S_{g}^{(1)}\rangle$
(normalized by the particles density) increases for larger $\phi$. This means that the generalized EP $S_{g}$ we proposed may partially recover
the information loss when simply treating the active particles system
as an effective system with high temperature by considering the interparticle
correlations at a coarse-grained level via Fox approximation. The difference between $\langle\Delta S_{g}\rangle$ and $\langle\Delta S_{g}^{(1)}\rangle$ becomes significant especially
in a high density system.

Furthermore, in Fig.\ref{fig:2}(b), we numerically calculate the
TUR parameter $\eta_{g}=\frac{2\langle\Theta\rangle^{2}}{Var[\Theta]\langle\Delta S_{g}\rangle}$
and $\eta_{g}^{(1)}=\frac{2\langle\Theta\rangle^{2}}{Var[\Theta]\langle\Delta S_{g}^{(1)}\rangle}$ for a large range of $\tau_{p}$ when $v_{0}=50$ and $v_{0}=100$.
If $\eta_{g}\leq1$, our TUR Eq.(\ref{eq:TUR-1}) is established,
otherwise, the TUR is invalid. The TUR still holds for a quite
large $\tau_{p}=10$, even though the Fox approximation
might break down for such a large $\tau_{p}$. Further increasing
the value of $\tau_{p}$, it can be observed that our TUR Eq.(\ref{eq:TUR-1})
also fails. In addition, we also plot the persistent time dependence
of the generalized EP for larger range of $\tau_{p}$ in Fig.\ref{fig:2}(c).
More interestingly, we find that $\Delta S_{g}$ has a non-monotonic
dependence on the persistence time, which is not consistent with the
physical expectation that increasing the persistent time displaces
the active particle systems progressively away from equilibrium. The non-monotonic persistence time dependence implies that the measure of dynamic irreversibility
is not monotonically related to the degree of departure from equilibrium
quantified by character of the ``effective equilibrium'' hypothesis
necessarily\cite{Flenner2020active}. In our opinion, this nontrivial phenomenon is mainly due to the fact that the coarse-grained method may not effectively restore the irreversibility of the system in large persistent-time regime. Actually, the establishment of the breakdown of the time-reversal symmetry for active systems with significant persistent motion is of great challenge. Thus, a deeper investigation of this disconnect is still a open question and deserves further investigation.

At last, we also validate the entropic bound
Eq.(\ref{eq:bound}) $\langle\dot{S_{g}}\rangle\leq\left\langle \bm{D}\cdot\left(\bm{F}^{eff}\right)^{2}\right\rangle $.
In Fig.\ref{fig:2}(d), we choose $\tau_{p}=0.0167$ and the entropic
bound has been numerically proved by varying the self-propelling velocity
$v_{0}$.

\subsection{Generalized entropy production and MIPS}

\begin{figure}
\begin{centering}
\includegraphics[width=1\columnwidth]{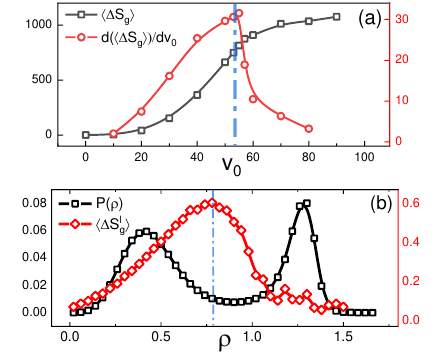}
\par\end{centering}
\caption{(a) The generalized entropy production $\langle\Delta S_{g}\rangle$
and the corresponding first derivative have been plotted over the
kinetic phases of MIPS. Here, the particles density $\phi=0.768$
to guarantee the occurrence of MIPS. Vertical blue dashed dot lines
indicate the critical velocity $v_{0}^{c}$. (b) Density and generalized
entropy production $\langle\Delta S_{g}\rangle$ in a local lattice
for ABPs model with MIPS have been plotted. Here, we choose $\phi=0.768$
and $v_{0}=80>v_{0}^{c}$. Vertical blue dashed dot lines indicate
the interface of MIPS.}

\label{fig:3}
\end{figure}

Another fundamental question in the context of stochastic thermodynamics
is whether entropy production/dynamical irreversibility can act as
a tool for typifying phase transitions. Insight into this question
has been gained in some recent studies \cite{Crochik2005PRE,Tome2012entropy,Shim2016macroscopic,Noa2019entropy,xiao2008entropy,xiao2009stochastic,seara2021irreversibility,cao2020design,Fodor2020dissipation,tociu2019how}.
For instance, by analyzing the majority-vote model, Noa $et$ $al.$\cite{Noa2019entropy}
have argued that there are specific hallmarks of entropy production for
a given transition, whether it is continuous or discontinuous.

\begin{figure}
\begin{centering}
\includegraphics[width=1\columnwidth]{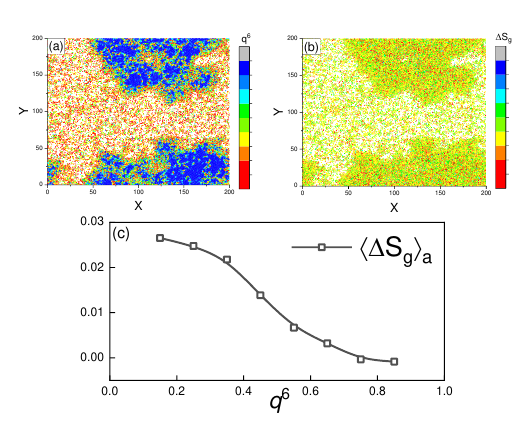}
\par\end{centering}
\caption{Snapshot configurations for $\phi=0.768$ and $v_{0}=80>v_{0}^{c}$
of the MIPS have been plotted: (a) local order parameter $q^{6}$,
(b) local entropy production $\langle\Delta S_{g}\rangle$ proposed
by us. In Fig. \ref{fig:4}(c), we show that the local entropy production
$\langle\Delta S_{g}\rangle$ decreases for higher $q^{6}$, which
means that the entropy production of the particles in defects is larger.}

\label{fig:4}
\end{figure}

We now focus on the relation between the dynamical irreversibility and
MIPS in ABPs system. The simulations are performed at a fixed particle
density $\phi=0.768$ by varying the velocity $v_{0}$, and the MIPS
occurs at a critical value $v_{0}^{c}|_{\phi=0.768}\approx54.3$.
In Fig.\ref{fig:3}(a), the steady-state generalized EPs, $\langle\Delta S_{g}\rangle$,
have been shown. As $v_{0}$ increases, the systems are driven far
away from equilibrium with $\langle\Delta S_{g}\rangle$ increasing.
Further, we find the $\langle\Delta S_{g}\rangle-v_{0}$ derivative
has inflection points and reach the maximum near the phase transition
point, providing the evidence that
the generalized EP $\Delta S_{g}$ can be used to indicate the large
scale MIPS. On the other hand, we also calculate the local entropy
production (density), $\langle\Delta S_{g}^{l}(\bm{x})\rangle$, by
averaging the particles' entropy production in a local lattice around
given position $\bm{x}$. As shown in Fig.\ref{fig:3}(b), the local
EP shows a strong contribution in the vicinity of the interfaces between
phases\cite{nardini2017entropy}. In addition, we find that local
entropy production almost vanishes in the high density phase due to
the dynamical arrest effect\cite{Nemoto2019optimizing,Cagnetta2017large}.
Thus, the coarse-grained entropy production $\Delta S_{g}$ also acts
as a reliable measure to determine the boundary of MIPS. Finally,
to investigate the connection between EP and defects, typical snapshots
for $\phi=0.768$ and $v_{0}=80>v_{0}^{c}$ of the MIPS have been
shown in Fig.\ref{fig:4}. Specifically, the structural information
coded in the spatial distribution of the local order parameter for
individual particles has been shown in Fig.\ref{fig:4}(a) with the
generalized EP in Fig.\ref{fig:4}(b) as a contrast. The defects are
found to allow for the increase in the generalized EP of active particles.
To confirm this conclusion clearly, we calculate the mean EP, $\langle\Delta S_{g}\rangle_{a}$,
by averaging $\Delta S_{g}$ for all particles with same $q^{6}$
in high density phase in Fig.\ref{fig:4}(c). Clearly, the mean EP
$\langle\Delta S_{g}\rangle_{a}$ decreases with the local order parameter
$q^{6}$ increasing as expected.

\section{Conclusions}

In conclusion, we focus on the thermodynamic quantities
for active systems over a finite time interval in steady states. We establish the stochastic thermodynamics for many-body active
particles system based on an approximate FPE obtained via a time-local approximation.
By mapping the systems into an equivalent Langevin equation, one can
identify a generalized trajectory-dependent EP $\Delta S_{g}$, wherein
particle activity comes into play by a configuration-dependent diffusion
coefficient and a many-body effective interaction force. The relationship
between the generalized EP and the rich collective behaviors of active
matter has been illustrated. Precisely, we utilize the generalized
entropy production to identify the phase transition point, the interface
and the defects in high density phases of MIPS, showing that the
generalized EP acts as a tool to quantify the dynamical
irreversibility on a macroscopic scale. Furthermore, the TUR and entropic bound for the currents
in the steady state can be established successfully for a large range of persistent time. In contrast, we show that simply mapping
the system to an equivalent one with an effective temperature does
not capture the right bounds, highlighting the suitable coarse-grained
approach via Fox approximation. Due to the link between TURs and anomalous diffusion\cite{Hartich2021unifying}, our results may help to bound the timescale of anomalous kinetics in active systems. We believe that our work can provide
a deeper understanding of the stochastic thermodynamics
in many-body active particles system.

\section*{Author Contributions}
Z.-Y.C. and J.S. contributed equally to this work.

\section*{Acknowledgments}
This work is supported by MOST(2018YFA0208702), NSFC (32090044, 21973085, 21833007, 21790350).

\section*{Author declarations}
The authors have no conflicts to disclose.

\section*{Data availability}
The data that support the findings of this study are available from the corresponding author upon reasonable request.

\bibliographystyle{apsrev} 

\expandafter\ifx\csname natexlab\endcsname\relax
\global\long\def\natexlab#1{#1}%
\fi \expandafter\ifx\csname bibnamefont\endcsname\relax
\global\long\def\bibnamefont#1{#1}%
\fi \expandafter\ifx\csname bibfnamefont\endcsname\relax
\global\long\def\bibfnamefont#1{#1}%
\fi \expandafter\ifx\csname citenamefont\endcsname\relax
\global\long\def\citenamefont#1{#1}%
\fi \expandafter\ifx\csname url\endcsname\relax
\global\long\def\url#1{\texttt{#1}}%
\fi \expandafter\ifx\csname urlprefix\endcsname\relax
\global\long\def\urlprefix{URL }%
\fi \providecommand{\bibinfo}[2]{#2} \providecommand{\eprint}[2][]{\url{#2}}

\end{document}